\documentstyle[twoside,fleqn,espcrc2]{article}
\input epsf.sty

\title{The broad-band X-ray spectrum of RE~0751+14 (PQ Gem)}

\author{T. Belloni\address{Astronomical Institute ``A. Pannekoek'',
	University of Amsterdam, Amsterdam, The Netherlands}
	\thanks{TB is supported by NWO under grant PGS 78--277}
       G. Matt\address{Dipartimento di Fisica, Universit\`a degli 
       studi ``Roma Tre'', Roma, Italy},
       D. De Martino\address{Osservatorio Astronomico di Capodimonte,
       Napoli, Italy},
       L. Chiappetti\address{IFCTR/CNR, Milano, Italy},
       F. Giovannelli\address{IAS/CNR, Frascati, Italy},
       F. Haberl\address{MPE, Garching, Germany},
       J. Osborne\address{University of Leicester, Leicester, UK},
       K. Mukai\address{NASA/GSFC, Greenbelt, USA},
       S. Molendi$^{\rm d}$,
       M. Mouchet\address{Observatoire de Paris-Meudon, Meudon, France}}

\begin{document}

\begin{abstract}

RE~0751+14 is a member of the class of Soft Intermediate Polars.
Unlike classical Intermediate Polars, they are characterized by the
presence of a soft component in their X-ray spectra, in addition to the
hard component typical of these systems.
The broad-band coverage of the Narrow-Field instruments on board 
{\it BeppoSAX} is ideal for the study of such objects. Here we report the
preliminary results of the analysis of a {\it BeppoSAX} pointing to RE~0751+14.
Both components are clearly detected, and their temperature determined
($\sim$50 eV and $\sim$40 keV), enabling the possibility of studying
them simultaneously.
\end{abstract}

\maketitle

\section{INTRODUCTION}

Intermediate Polars (IPs) are a subclass of cataclysmic variables in which
the magnetic field of the white dwarf strongly influences the accretion
flow, channeling the material onto the magnetic poles. Contrary to the class
of Polars, these systems are not synchronized, i.e. the white dwarf rotation
period is shorter than the orbital period of the binary. An accretion
disk might be present, but disrupted by the magnetic field
before reaching the white dwarf surface. As the accreting gas reaches the
white dwarf, a strong shock is formed, resulting in the emission of X-rays
(see \cite{Patt94} for a review).

Differently from Polars, IPs do not generally show detectable polarized
radiation in the optical/IR bands, suggesting that they possess
lower magnetic fields, although the detection of polarized radiation in a
few systems has raised the question of whether IPs have similar magnetic
fields to Polars but are still asynchronized because of their longer
orbital periods \cite{Chan92}.

Coherent X-ray pulsations, ranging from a few tens to a few thousands of
seconds, are a signature of channeled accretion towards the magnetic poles
of the white dwarf. While in the optical this periodicity $P_\omega$ appears
as a large-amplitude quasi-sinusoidal modulation, in the X-rays it is quite
complex \cite{Nort95}. The X-ray spin pulse amplitude is strongly energy
dependent, being greater ($\sim$50\%) at low energies. This suggests that
the modulation is at least partly due to phase-dependent photoelectric
absorption. Besides the spin, variations at the orbital $P_\Omega$ ($\sim$
hours) and beat $P_{\omega-\Omega}$ periods are also observed.

\begin{figure*}[htb]
\epsfysize=11cm
\vspace{9pt}
\hspace{1.0cm}\epsfbox{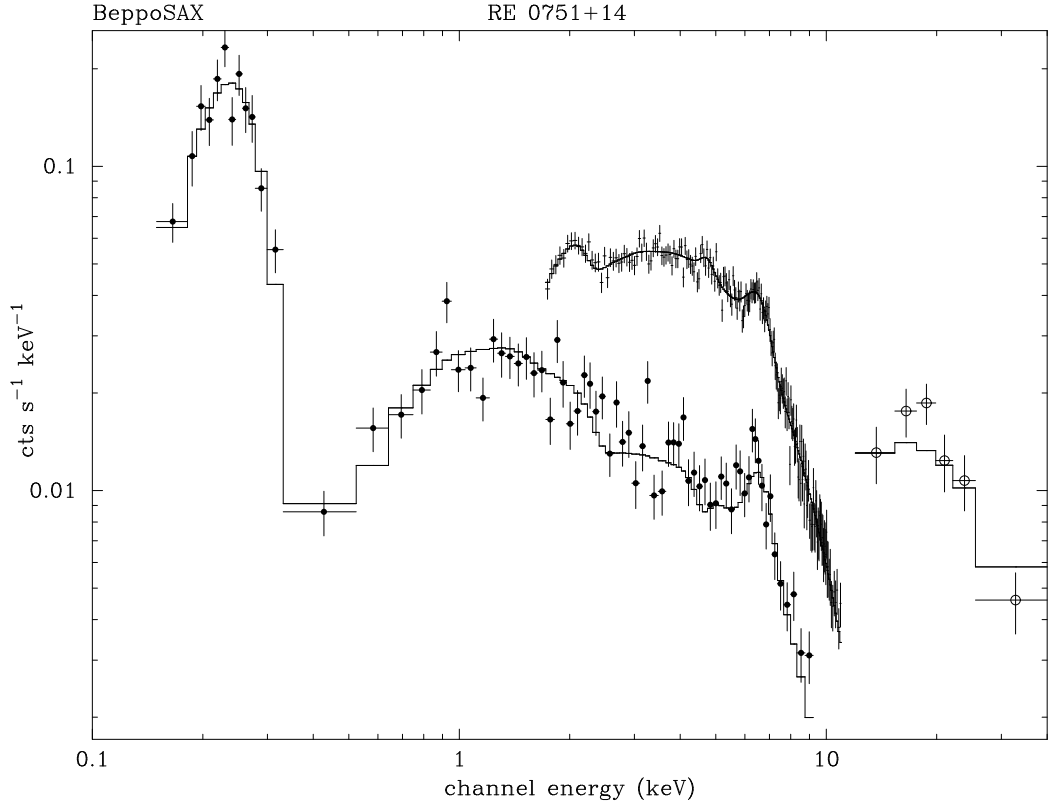}%
\caption{Broad-band X-ray spectrum of RE~0751+14 (LECS/MECS/PDS). The line
       is the best fit model described in the text.}
\label{fig:1}
\end{figure*}

IP's are harder than polars in that they generally do not have visible
soft X-ray components, which could be consistent with the idea that high
field magnetic accreting white dwarfs tend to suppress hard X-ray 
emission \cite{Beue95}. On the other hand, the possible
presence of large additional absorption is likely to make such a
component unobservable. Observations made with the {\it Ginga} satellite
showed that for some systems a good fit is obtained with a double
bremsstrahlung model in which the two components have the same temperature
but different intrinsic absorption \cite{Nort89,Ishi91}.
An obvious interpretation for this is that the absorbing material is
spatially or temporally inhomogeneous.

Recently, an substantial soft X-ray component has been discovered in a few
sources of the IP class. Four of these sources were discovered with {\it ROSAT}
\cite{Maso92,Habe95,Burw96}. Based on the
similarity of their energy distribution with that of Polars, and on
evolutionary considerations \cite{Hame87}, it is thought that 
these soft IPs are the actual progenitors of Polars.
Indeed, their soft component has been modeled as a blackbody with temperatures
between 40 and 60 eV, slightly higher than the typical values for Polars.
In the two systems which were bright enough to allow a more detailed
spectral analysis, an additional weaker hard component has been observed
\cite{Habe94}.

\begin{table}[hbt]
\setlength{\tabcolsep}{1.5pc}
\newlength{\digitwidth} \settowidth{\digitwidth}{\rm 0}
\catcode`?=\active \def?{\kern\digitwidth}
\caption{Preliminary best fit parameters corresponding to the model described
        in the text.}
\label{tab:1}
\begin{flushleft}
\begin{tabular}{@{}l@{\extracolsep{\fill}}lc}
\hline
Parameter~~~~~~~~~ & Best fit value\\
\hline
N$_H$            & $1.1\times 10^{20}$cm$^{-2}$\\
\hline
kT$_{bb}$        & 51$\pm$11 eV\\
R$_{bb}$         & 16 (d/260 pc) km\\
L$^{bol}_{bb}$   & 2.5$\times 10^{32}$ (d/260 pc)$^2$ erg/s\\
\hline
kT$_{br}$        & 43$\pm$13 keV\\
L$_{br}$         & 5.8$\times 10^{31}$ (d/260 pc)$^2$ erg/s\\
\hline
E$_{line}$       & 6.62 keV\\
$\sigma_{line}$  & 0.38 keV\\
\hline
N$_H$            & 7.6$\times 10^{22}$ cm$^{-2}$\\
cov. frac.       & 0.46\\
\hline
\end{tabular}
\end{flushleft}
\end{table}

It is thus crucial to properly determine the energy distribution at low
and high energies in order to evaluate the influence of the magnetic field,
of the accretion rate and of the inferred accretion geometry on the
existence of a strong soft component. With this goal in mind, we have 
started a campaign to observe three of these systems (RE~0751+14,
RX~J0558.0+5333 and FO~Aqr) with {\it BeppoSAX}: the unique broad band
coverage of the Narrow Field Instruments on board {\it BeppoSAX} is ideal
for the study outlined above.

\section{RE~0751+14}

RE~0751+14 (PQ Gem) has been discovered as a bright EUV source in the
{\it ROSAT} WFC All-sky survey \cite{Maso92}. It is the second IP 
ever detected in the EUV (the first being EX~Hya) and the first 
EUV-selected. In common with classical IPs the system has an
asynchronously rotating white dwarf ($P_\omega$=13.9 min), much
shorter than the orbital period $P_\Omega$=5.2 hr
\cite{Maso92,Rose93,Hell94}, a strong hard
X-ray modulation at $P_\omega$ \cite{Maso92,Duck94}
and optical modulation at the beat frequency \cite{Maso92}. 
The spin-modulated polarization and a photometric orbital
variation in the red \cite{Rose93,Piir93,Pott97},
and of course the presence of a strong
soft X-ray component modulated at $P_\omega$ are features which link
RE~0751+14 to Polars, indicating a magnetic field stronger than 
classical IPs.
Recently, Mason \cite{Maso97} reported large variations of $P_\omega$. The
white dwarf spins down with $\dot P = 1.1\times 10^{-10}$s/s, the
largest spin variation observed in an IP.

Its X-ray spectrum, as observed in two separate observations 
with {\it ROSAT} and
{\it Ginga} has been modeled with a soft blackbody (kT=46 eV) and a 
partially-absorbed hard bremsstrahlung (kT$\sim$20 keV). Additionally,
a narrow iron line at 6.7 keV is also required to fit the data \cite{Duck94}.
No time variations in the parameters of the partial-covering model were detected 
within the errors. A
phase-resolved spectral analysis of both the soft and the hard components
has also been performed, but the
non-simultaneity of the data makes the results uncertain \cite{Duck94}.

\section{BEPPOSAX OBSERVATIONS}

We observed RE~0751+14 with the Narrow Field Instruments on board {\it BeppoSAX}
in 1996 between Nov. 9th 16:48 UT and Nov 12th 6:24 UT. The total exposure
time is 105 ksec for MECS, HPGSPC and PDS, but only 24 ksec for the LECS,
because of additional observational constraints. For a description of
the satellite and its instruments see \cite{Boe97a,Parm97,Boe97b,Fron97}.

The source has been detected in all four instruments. Our preliminary
analysis is limited to LECS, MECS and PDS, ensuring a coverage between
0.1 and 200 keV with a gap between 10 and 14 keV.

Data preparation and linearization was performed using
the {\sc Saxdas v1.0} package under {\sc Ftools} environment.
For LECS and MECS, source photons have been extracted from a 
circle centered on the source with radius 4$'$. 
The standard background has been used for background subtraction.

\section{THE BROAD-BAND SPECTRUM}

The 0.1--200 keV spectrum of RE~0751+14 obtained combining all three
instruments is shown in Figure 1. The source is clearly detected from
0.2 to 40 keV. The net count rates in the different instruments are
0.13 cts/s (LECS), 0.30 cts/s (MECS, 3 units) and 0.26 cts/s (PDS).
In order to fit the data we adopted a model consisting of interstellar
absorption, a soft blackbody, a hard bremsstrahlung, a gaussian line
around 6 keV and a partial covering absorption model (see \cite{Duck94}).
In order to account for differences in the cross-calibration, the 
normalizations for the single instruments have been left free and 
corrected for calibration effects after the fit. The best fit parameters
are shown in Table 1 (reduced $\chi^2\sim$1.0). Given the preliminary
status of the analysis, only the 1$\sigma$ errors on the two temperatures
are reported.

Thanks to the high-energy response provided by the PDS, it is for the 
first time possible to determine the temperature of both the hard and
soft components simultaneously. Notice that they are roughly three
orders of magnitude apart. The values are compatible with the
results of \cite{Duck94}. 
The ratio of the soft to hard flux is $\sim$4.3, also consistent with
what found by \cite{Duck94}. At variance with what observed with
Ginga, the iron line is rather broad and centered at lower energies.
This is likely to be the result of the blending of multiple narrow
lines such as seen in the IP AO~Psc by \cite{Hell96}.
The derived size of the emission area for the blackbody component is
consistent with what derived by \cite{Habe95}, corresponding
to a fractional area of the white dwarf $f\sim 1\times 10^{-5}$.
\section{DISCUSSION}

These results show that the soft X-ray emission in PQ Gem originates
from a very small area which could be related to the footprints on the
white dwarf surface of an arc-shaped accretion curtain \cite{Rose88}.
On the other hand, the blackbody temperature derived here, consistent with
the {\it ROSAT} results, falls into the high value range for Polars, 
while the thermal plasma temperature is close to the standard IPs.
Interestingly, the derived soft-to-hard X-ray flux ratio is more appropriate
to a Polar system.

We have compared the observed flux ratio (computed in the {\it ROSAT} band
0.1--2.4 keV) to those observed in Polar systems (see Fig 8. in \cite{Beue97}).
From this we can estimate a magnetic field strength of $\sim$10 G.
This value is compatible with the estimates by \cite{Pott97}.

\end{document}